# Identification of an Open-loop Plasma Vertical Position Using Fractional Order Dynamic Neural Network

**Z. Aslipour[1] and A. Yazdizadeh**

*Department of Electrical Engineering, Shahid Beheshti University (SBU),*
*Velenjak, Tehran, Iran*
*E-mail: z_aslipour@sbu.ac.ir*

ABSTRACT: In order to identify complicated systems, more prominent and promising methods are needed among which we may refer to fractional order differential equations. The aim of this paper is to propose a fractional order nonlinear model to predict the vertical position of a plasma column system in a Tokamak by using real data from Damavand Tokamak. The system is identified based on a newly introduced fractional order dynamic neural network. The proposed fractional order dynamic neural network (FODNN) is an extension of the integer order dynamic neural network that employs the so called fractional-order operators. Due to the rich structure of the proposed neural network, it models the complicated systems with less error.

Training rule is derived based on a Lyapunov-like analysis that ensures a bound for the “identification error” and tends towards zero as time leans towards infinity. FODNN is implemented and comparison of the numerical simulation results with experimental results shows that performance of the proposed method by using fractional order neural network is preferred to the integer neural network.

[1] Corresponding author.

## Contents

## 1. Introduction

There are many reasons for developing a mathematical model for an actual plant, among which we may refer to the necessity of designing an identifier in all model-based controller schemes [1]. In dynamic systems, integer order differentiation and partial order differentiation can be used to describe the behaviour of the system. In fractional order differential equations, there is an additional parameter, namely, fractional order, which is important to be selected properly to get better results while modelling and simulating the plant. The fractional order derivative operator is part of the fractional calculus that was introduced 300 years ago. The main advantage of the fractional derivative, in comparison with integer order methods, is that it provides an excellent tool for describing memory and hereditary properties of the processes. This advantage cannot be ignored and as a result, fractional modelling has been employed in different processes [2-5].

Analytical modelling is basically development of a mathematical relationship between input and output based on the physical laws or based on some numerical phenomena. Neural network techniques seem to be very useful to identify different classes of nonlinear systems. Major properties that motivate the use of neural networks are as follows: their nonlinear characteristics make them appropriate for identifying nonlinear systems and their learning characteristics are ideal for adapting to different environmental conditions [6]. Neural networks can be categorized as static (Feedforward) and dynamic (recurrent or differential) neural networks. The main disadvantages of the static structure are as follows: a) slow learning rate, b) lack of possessing memory, so their outputs are uniquely dependent on the current inputs. In contrast, dynamic neural networks can successfully overcome these shortcomings and demonstrate promising behaviour in the presence of un-modelled dynamics because their structure includes an intrinsic feedback. So far, different structures for the integer order dynamic neural network have been used for different applications [7-11].

The use of fractional-order models in system identification was initiated in the late 1990s and the beginning of this century. Several techniques are available for identification of the systems using fractional order derivatives, including equation error and output error approaches [12-15]. Nonlinear fractional order system identification using the static neural network is studied in [16-18]. In recent years, different types of the fractional order operators have been engaged in dynamic neural networks. Formulation and evaluation of the dynamic behaviour of FODNN have been studied by many researchers [19-24]. In [25, 26] a FODNN is used to identify linear and nonlinear integer order system.

The Damavand Tokamak is a small size research machine for fusion-related studies like plasma diagnostics [27, 28]. In Damavand Tokamak, hydrogen plasma is sustained for 21ms with a plasma current peak of about 35 kA. Several methods have been introduced for identification of the Damavand Tokamak system. H. Rasouli and N. Darestani have provided a nonlinear model based on a static neural network with delay lines [29, 30]. In addition, a linear model has been used around the defined operating point in [31]. Since the fractional operators are usually able to model more complicated systems, this physical phenomenon can be stated by fractional-order models. The use of fractional calculation has also been reported for modelling of the chaotic behaviour of plasma and magneto hydrodynamic (MHD) instability of plasma in Tokamaks [32]. The Damavand Tokamak, has been identified by using a linear fractional order model in [33, 34] as well.

This paper is motivated by the need to create a nonlinear subspace model which is suitable for controller design. In this work, we propose a FODNN identification technique to model the Damavand Tokamak. It should be noted that the plasma-coil system is a Multi Input Single Output (MISO) system.

The rest of the paper is organized as follows: In section 2, Damavand Tokamak and the vertical position control of the plasma column is described. In section 3, the FODNN for identifying the nonlinear system is proposed and discussed in details. The identification procedure uses collected data from series of identification experiments. It is explained why this tool is selected to identify the Damavand Tokamak and then its structure and learning laws are introduced. In section 4, the proposed identification model is compared to experimental results in order to show the accuracy of the model. Finally, conclusions are given in section 5.

## 2. Description of Damavand Tokamak and Experimental Setup

Among the small Tokamaks, the Damavand Tokamak is most important due to its large elongated plasma cross-section and having an active control system for the movement and shape of the plasma as well as various other diagnostic devices. Unfortunately, this elongation leads to instability of the plasma vertical position. Indeed, to stabilize the plasma position, an active feedback system is needed. Plasma is formed in the centre of the chamber with R= 36 cm with a maximum current of 35 KA in 21 ms [38]. In the Damavand Tokamak, due to the inherent instability of plasma vertical position, a suitable method for identification based on the experimental data is the direct closed loop method. In this method, the process is controlled in a closed loop structure while the input and output data of the process are used for identification of the system. A block diagram showing the plasma vertical position closed loop control system and direct model are illustrated in Figure 1.

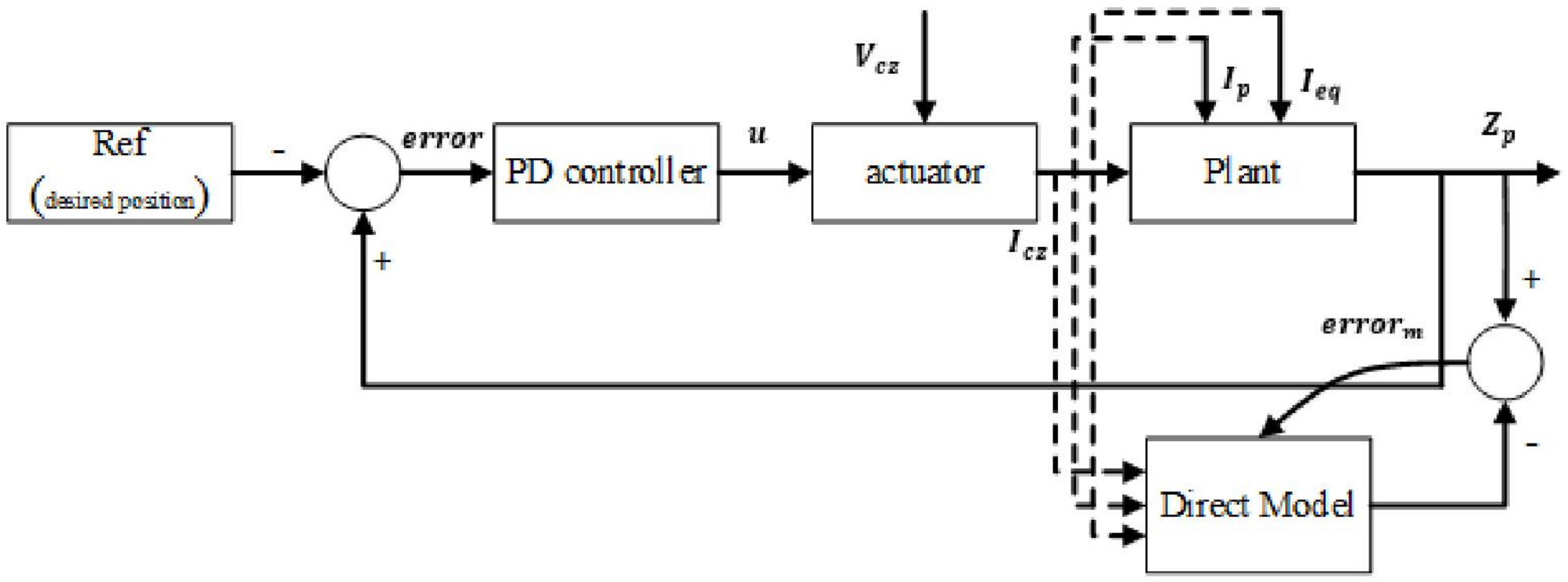

Figure.1. Block diagram of plasma radial motion closed loop control system in Damavand Tokamak

As shown in Figure 1, this block diagram consists of a reference signal generator circuits, PD controller circuits, actuator circuits and Tokamak system. The Proportional-Derivative (PD) controller transfer function, actuator circuit model and main parameters of the Damavand Tokamak are given in [38] in details. The parameters are kept constant in all tests. Considering the total time of plasma electrical discharge (shot) that occurs in a duration of 21ms, online identification is hard. In order to correct biases and to select the appropriate time interval of data in each shot, a data pre-processing step is required after which the implemented data for identification purpose is shown in Figure 2 in a training phase.

In the next part, the nonlinear dynamic model for the Damavand Tokamak is identified. To achieve this several shots with fixed z-position have been gathered, the identification process has taken place for a single shot as training data and finally it has been validated for others. More descriptions about the method and the results are presented in the subsequent sections.

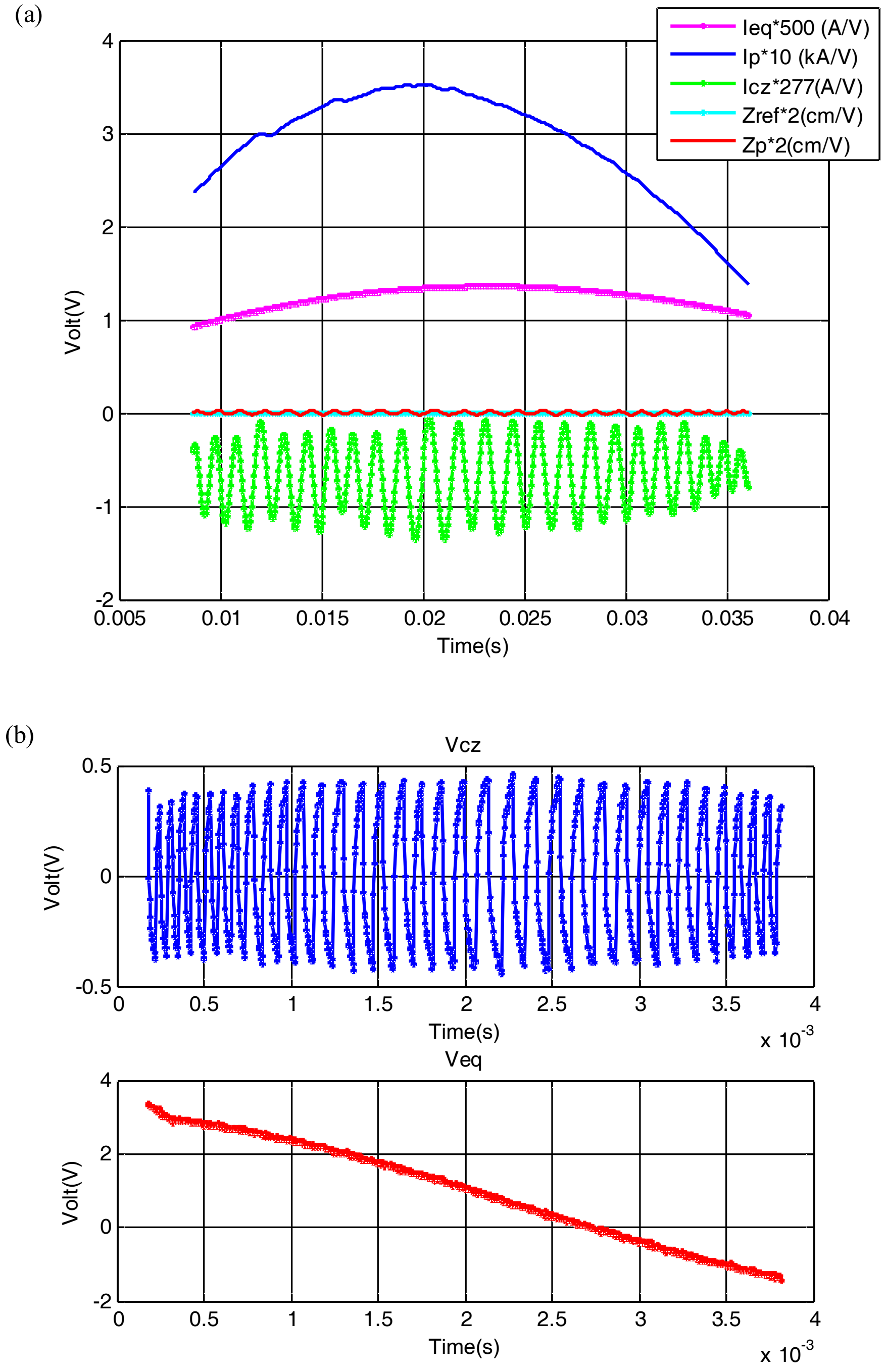

Figure 2. Experimental data for training and validation of identified model (a) Equilibrium coil current , Plasma current, Vertical position control coil current for production of elongated plasma with Zp=0, Plasma vertical position, Plasma vertical position reference. (b) Vertical position control coil voltage, Equilibrium coil voltage for production of elongated plasma with Zp=0 (shot (2015/27/05 Shot# 83)).

## 3. Identifier Design by Using Fractional Order Dynamic Neural Network (FODNN)

Different methods have been developed for linear and nonlinear systems identification. Before selection of the method, linearity or nonlinearity of the system behaviour must be determined. This subject has been discussed in [31]. It is shown that the Damavand Tokamak has a nonlinear inherent. Therefore, in this paper a nonlinear system identification method for obtaining a complete model of Damavand Tokamak is used. Undoubtedly, neural networks are effective tools for identification of such systems due to their unique features. The Damavand Tokamak has been identified by neural networks in different works [35, 36]. The use of fractional calculation is considered more appropriate in modelling the chaotic behaviour of plasma and magneto hydrodynamic (MHD) instability of plasma in Tokamak [32]. In the last researches on Damavand Tokamak, a linear fractional order model as transfer function has been identified [33, 34]. Since the plasma behavior has a fractional nature and Damavand Tokamak system is nonlinear, in this paper, we focus on a fractional order nonlinear method (FODNN) for identification of the Damavand Tokamak system. Finally, a fractional order nonlinear model as a state-space representation is given for plasma vertical position, based on the FODNN identification structure that is explained in this section.

In the proposed method, it is assumed that the states of the system are available, which is the case in the Damavand Tokamak. In addition, the number of FODNN states are selected as the number of the states of the system. By using a Lyapunov-like analysis, FODNN parameters are adapted so that the identification error is stable[25].

The nonlinear system to be identified is given as:

$$\dot{x}_t = f(x_t, u_t, t) \quad , \quad x_t \in R^n \quad , \quad u_t \in R^m \tag{3}$$

We consider the following FODNN for identification:

$$D^q \hat{x}_t = A\hat{x}_t + W_{1t}\sigma(\hat{x}_t) + W_{2t}\emptyset(\hat{x}_t)\gamma(u_t) \tag{4}$$

Fractional order calculus has three common definitions of differential operators: Riemann–Liouville, Caputo definitions and Grunwald–Letnikov [37]. In the proposed method, the Caputo fractional order derivative is used, because its initial conditions are identical to the ones of integer order derivatives, which is well-understood in our physical situations and more applicable to real world problems.

The Caputo derivative of the fractional order of function $x(t)$ is defined as follows:

$$ {}_0^C D_t^q x(t) = D_t^{q-p} \frac{d^p}{dt^p} x(t) \tag{5}$$

$$= \begin{cases} \dfrac{1}{\Gamma(p-q)} \displaystyle\int_a^t \frac{x^p(\tau)}{(t-\tau)^{q-p+1}}\, d\tau & p-1 < q < p \in Z^+ \\ \dfrac{d^p x(t)}{dt^p} & q = p \end{cases}$$

Where $\Gamma(.)$ is the gamma function defined as:

$$\Gamma(\mathrm{z}) = \int_0^\infty e^{-t} t^{z-1} dt \tag{6}$$

In (4), $D^q$ denotes Caputo fractional-order derivative of order $q$ $(0 < q < 1)$, $\hat{x}_t \in R^n$ is the state of the fractional neural network and , $u_t \in R^m$ is its input. The matrix $A \in R^{n\times n}$ is a stable diagonal fixed matrix and $W_{1t} \in R^{n\times n}$ and $W_{2t} \in R^{n\times n}$ are the weights of the FODNN identifier. The vector function $\sigma(.) \in R^n$ is assumed to be n-dimensional with the elements increasing monotonically. The matrix function $\emptyset(.) \in R^{n\times m}$ is assumed to be diagonal.

$$\emptyset(\hat{x}_t) = diag(\emptyset_1(\hat{x}_t), \dots, \emptyset_n(\hat{x}_t)) \tag{7}$$

Function $\gamma(.) \in R^m$ is selected as $\|\gamma(u_t)\|^2 \le \bar{u}$. The typical presentation of the elements $\sigma_i(.)$ and $\emptyset_i(.)$ are sigmoid functions that satisfy 'the sector conditions' (see Figure 3)

$$\sigma_i(x_i) = \frac{a_i}{1 + e^{-b_i x_i}} - c_i \tag{8}$$

Since $\sigma(.)$ and $\emptyset(.)$ are chosen as sigmoid functions, they clearly fulfil the following Lipshitz condition:

$$\tilde{\sigma}^T \Lambda_1 \tilde{\sigma} \le \Delta_t^T D_\sigma \Delta_t \tag{9}$$

$$\gamma^T(u_t)\tilde{\emptyset}_t^T \Lambda_2 \tilde{\emptyset}_t \gamma(u_t) \le \bar{u}\Delta_t^T D_\emptyset \Delta_t$$

Where $\Lambda_1, \Lambda_2$, $D_\sigma$ and $D_\emptyset$ are known positive constants and $\tilde{\sigma}$, $\tilde{\emptyset}$ are defined as below:

$$\tilde{\sigma} = \sigma(\hat{x}_t) - \sigma(x_t) \tag{10}$$

$$\tilde{\emptyset} = \emptyset(\hat{x}_t) - \emptyset(x_t)$$

The structure of this FODNN is shown in Figure 4. The integrator operator $\frac{1}{s^q}$ is a fractional order operator.

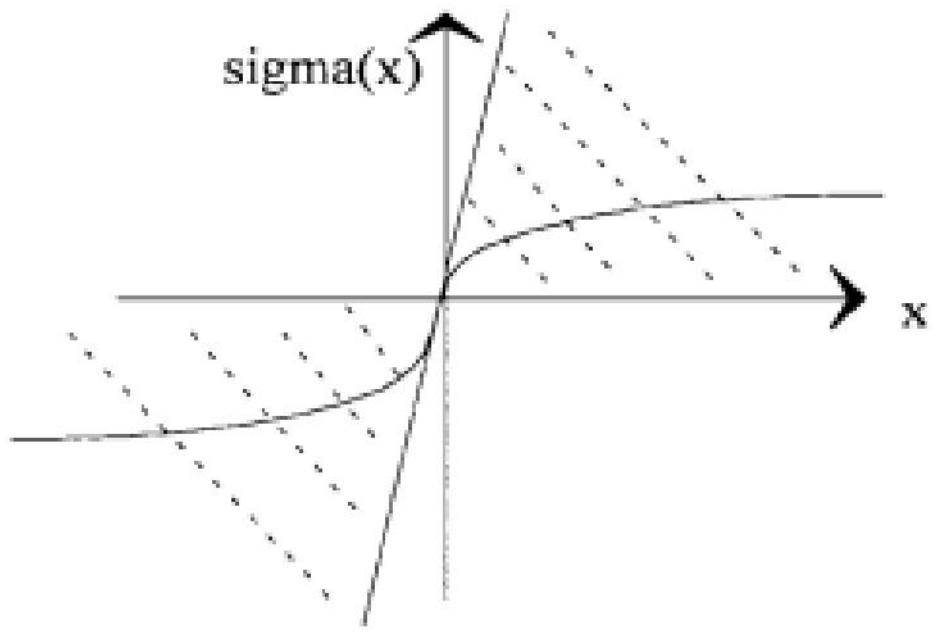

Figure 3: The shaded part satisfies "the sector condition".

Generally, FODNN (4) cannot follow the nonlinear system (3) exactly. The nonlinear system may be written as:

$$\dot{x}_t = Ax_t + W_1{}^*\sigma(x_t) + W_2{}^*\emptyset(x_t)\gamma(u_t) \tag{11}$$

where $W_1{}^*$ and $W_2{}^*$ are bounded unknown matrices as:

$$W_1{}^*\Lambda_1{}^{-1}W_1{}^{*T} \le \bar{W}_1 \tag{12}$$

$$W_2{}^*\Lambda_2{}^{-1}W_2{}^{*T} \le \bar{W}_2$$

where $\bar{W}_1$,$\bar{W}_2$, $\Lambda_1$ and $\Lambda_2$ are already known matrices.

Now to develop the training law, let us define identification error as:

$$\mathrm{e}_t = D^{q-1}\hat{x}_t - x_t \tag{13}$$

where $\hat{x}_t$ is the fractional order dynamic neural network identifier state and $x_t$ is the nonlinear system real state. The error dynamics is obtained as:

$$\begin{aligned} \dot{e}_t &= A\hat{x}_t + W_{1t}\sigma(\hat{x}_t) + W_{2t}\emptyset(\hat{x}_t)\gamma(u_t) - Ax_t - W_1{}^*\sigma(x_t) \\ &\quad - W_2{}^*\emptyset(x_t)\gamma(u_t) \\ &= A\mathrm{e}_t + W_1{}^*\tilde{\sigma} + W_2{}^*\tilde{\emptyset}\gamma(u_t) + \tilde{W}_{1t}\sigma(\hat{x}_t) + \tilde{W}_{2t}\emptyset(\hat{x}_t)\gamma(u_t) \end{aligned} \tag{14}$$

where $\tilde{W}_{1t}$ and $\tilde{W}_{2t}$ are defined as below:

$$\widetilde{W}_{1t} = W_{1t} - W_1{}^*$$
$$\widetilde{W}_{2t} = W_{2t} - W_2{}^* \tag{15}$$

A Lyapunov function is selected as:

$$V_t = \mathrm{e}_t{}^T P \mathrm{e}_t + tr\left(\widetilde{W}_{1t}{}^T k_1{}^{-1} \widetilde{W}_{1t}\right) + tr\left(\widetilde{W}_{2t}{}^T k_2{}^{-1} \widetilde{W}_{2t}\right) \tag{16}$$

Where $P \in R^{n\times n}$ is a positive definite matrix. According to (13), the derivative is:

$$\dot{V}_t = 2(e_t{}^T P \dot{\mathrm{e}}_t + tr\left(\dot{\widetilde{W}}_{1t}{}^T k_1{}^{-1} \widetilde{W}_{1t}\right) + tr\left(\dot{\widetilde{W}}_{2t}{}^T k_2{}^{-1} \widetilde{W}_{2t}\right)) \tag{17}$$

By substituting (14) into (17). We get:

$$e_t{}^T P \dot{\mathrm{e}}_t = \mathrm{e}_t{}^T P\left(\widetilde{W}_{1t}\sigma(\hat{x}_t) + \widetilde{W}_{2t}\emptyset(\hat{x}_t)\gamma(u_t)\right) + e_t{}^T P A \mathrm{e}_t + e_t{}^T P\left(W_1{}^*\tilde{\sigma} + W_2{}^*\tilde{\emptyset}\gamma(u_t)\right) \tag{18}$$

Using of simple operations in (17) may be concluded as:

$$\dot{V}_t \le e_t{}^T(PA + A^T P + P(\overline{W}_1 + \overline{W}_2)P + D_\sigma + D_\emptyset \bar{u} + Q_0)e_t - e_t{}^T Q_0 e_t + 2tr\left(\left(\dot{\widetilde{W}}_{1t}{}^T k_1{}^{-1} + P e_t \sigma(\hat{x}_t)^T\right)\widetilde{W}_{1t}\right) + 2tr\left(\left(\dot{\widetilde{W}}_{2t}{}^T k_2{}^{-1} + P\tilde{x}_t\gamma(u_t)^T\emptyset(\hat{x}_t)^T\right)\widetilde{W}_{2t}\right) \tag{19}$$

If we define:

$$R = \overline{W}_1 + \overline{W}_2$$
$$Q = D_\sigma + D_\emptyset \bar{u} + Q_0 \tag{20}$$

And the matrices $A$ and $Q_0$ are selected to satisfy the following conditions:

1) The pair$(A, R^{\frac{1}{2}})$ is controllable, the pair $(A, Q^{\frac{1}{2}})$ is observable;
2) Local frequency condition satisfies

$$A^T R^{-1} A - Q \ge \frac{1}{4}(A^T R^{-1} - R^{-1}A)R(A^T R^{-1} - R^{-1}A)^T \tag{21}$$

Then the following assumption can be established:

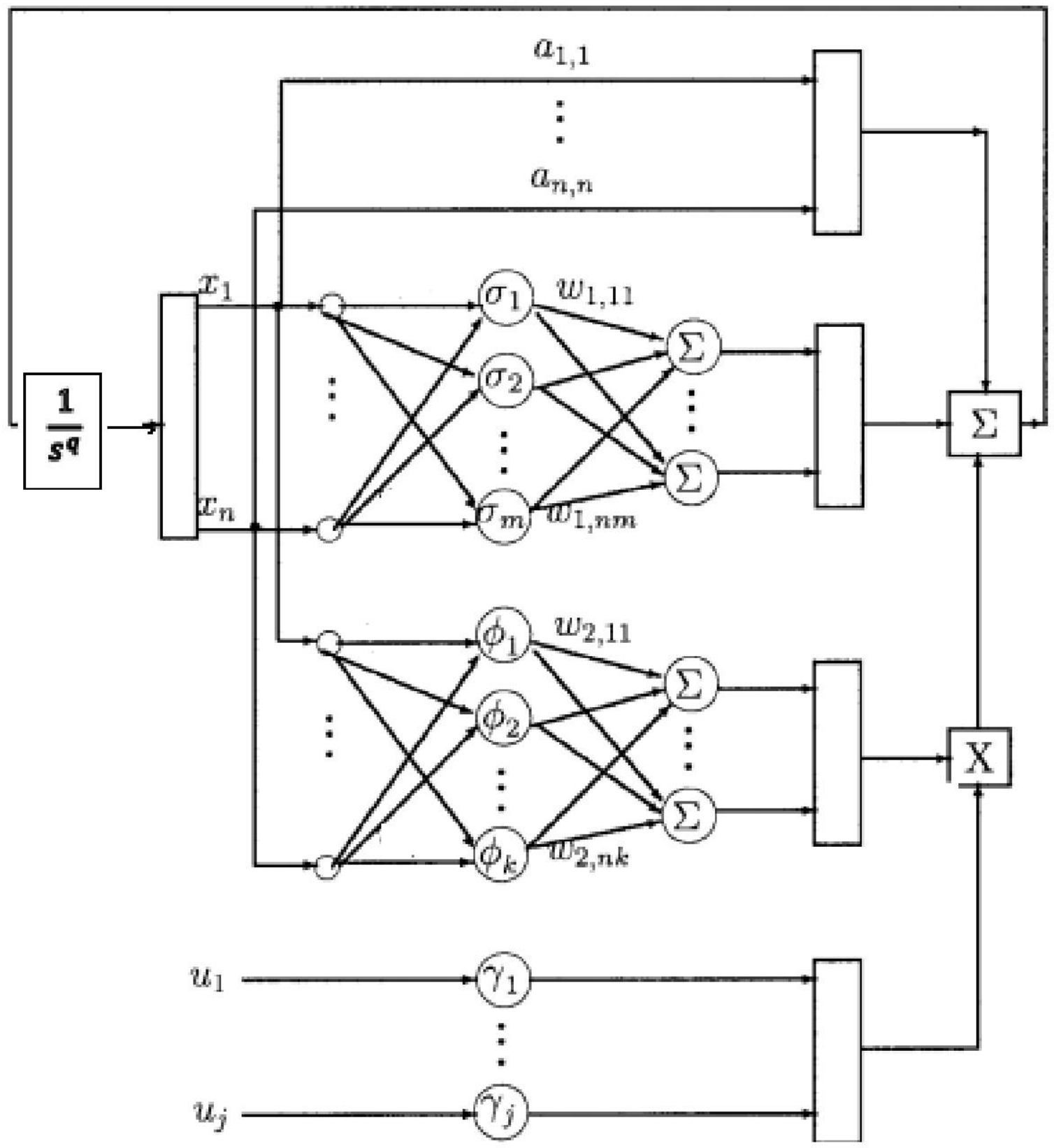

Figure 4: The general structure of the dynamic FODNN.

A1: There exist a stable matrix and a strictly positive definite matrix $Q_0$ such that the matrix Riccati equation:

$$PA + A^T P + PRP + Q = 0 \quad (22)$$

has a positive solution $P = P^T > 0$

This condition is easily fulfilled if we select $A$ as stable diagonal matrix. If the weights $W_{1t}$ and $W_{2t}$ are updated as:

$$\dot{W}_{1t} = -k_1 P e_t \sigma(\hat{x}_t)^T \quad (23)$$
$$\dot{W}_{2t} = -k_2 P e_t \gamma(u_t)^T \emptyset(\hat{x}_t)^T$$

where $P$ is the solution of the Riccati equation (22) and $k_1$, $k_2$ are the positive definite matrices, then the identification error (13), $W_{1t}$ and $W_{2t}$ are bounded and $\dot{V}_t \leq -e_t{}^T Q_0 e_t \leq 0$.

## 4. Validation of the Proposed Method by Using Experimental Data and Integer Order Model

The governing nonlinear dynamic equation of the Damavand Tokamak is given in [38]. Plasma vertical movement equations are given in the time domain as follows:

$$\dot{Z}_p - \delta_{cz}\dot{I_{cz}} - \delta_{eq}\dot{I_{eq}} = 0 \quad (24)$$

$$\alpha_{cz}\dot{Z}_p + \dot{I_{cz}} + \gamma_{cz}I_{cz} + \beta_{cz}\dot{I_{eq}} = \frac{V_{cz}}{L_{cz}} \quad (25)$$

$$\alpha_{eq}\dot{Z}_p + \dot{I}_{eq} + \gamma_{eq}I_{eq} + \beta_{eq}\dot{I}_{cz} = \frac{V_{eq}}{L_{eq}} \quad (26)$$

Where $V_{cz}$, $V_{eq}$, $I_{cz}$ and $I_{eq}$ are voltage on the vertical control coil, voltage on the equilibrium coil, vertical control coil current and equilibrium coil current respectively. $Z_p$ is the plasma vertical position. All signals mentioned in the above equations except $Z_p$ are divided by $I_p$(plasma current). Other coefficients are related to the physical and electrical characteristics of system that have been discussed in detail in [38]. To solve this system of equations, given that $I_p$, $V_{cz}$ and $V_{eq}$ are measured in any Tokamak discharge, the$Z_p$, $I_{cz}$ and $I_{eq}$ can be achieved completely by knowing the initial values.

Time variant differential equations (24), (25) and (26) can be written in this form:

$$\dot{x}(t) = f(x(t), u(t), t) \quad (27)$$

In these equations, $u(t) = \begin{bmatrix} 0 \\ \frac{V_{cz}(t)}{I_p} \\ \frac{V_{eq}(t)}{I_p} \end{bmatrix}$ is input vector to the system and $x(t) = \begin{bmatrix} Z_p(t) \\ \frac{I_{cz}(t)}{I_p} \\ \frac{I_{eq}(t)}{I_p} \end{bmatrix}$ is system states vector. The equation (27) is the general form of the nonlinear state equation of the Damavand Tokamak system.

In [29] authors have mentioned that the sampling time and delay between inputs and output of the Damavand Tokamak are effective in identification performance. In this paper, we selected the delay between inputs ($I_{cz}$ , $I_{eq}$) and output ($Z_p$) equal to 240 µs and sampling time equals to 40 µs. The above mentioned amounts have been achieved based on different trials.

In this section, to show the capabilities and performance of the proposed method, a fractional order and an integer order model based on the dynamic neural network are tested. In the proposed neural network, activation functions are selected as sigmoid functions:

$$\sigma(x_i) = \frac{35}{1 + e^{-35x_i}} - 0.5$$

$$\emptyset(x_i) = \frac{10}{1 + e^{-10x_i}} - 0.05$$

The main parameters of the neural network identifier (15) are selected by a 'try to test' method as:

$$A = \begin{bmatrix} -100 & 0 & 0 \\ 0 & -100 & 0 \\ 0 & 0 & -100 \end{bmatrix} \qquad k_1 = k_2 = 10^3 I$$

It is obvious that the proposed parameters must satisfy the mentioned assumptions in section 3 to guarantee that the identification error remains bounded. The conditions are assumed to be identical for both fractional order and integer order dynamic neural networks so that we can properly compare the models. The additional parameter of the fractional order model identifier which is given by (4) is $q$. There is no systematic procedure for obtaining fractional order selection. Therefore, it is chosen by trial and error, so that convergence of the FODNN is guaranteed and error index is minimized. This parameter for Damavand Tokamak identification is assumed to be equal to 0.7 and in the integer neural network model is equal to 1. The numerical Predictor-Corrector (PC) algorithm is used to solve the fractional order differential equation (4).

The Damavand Tokamak identification via FODNN has been performed using Matlab/Simulink by direct programming of relations. In the first step, the FODNN is constructed as in (4). Two sigmoid functions $\sigma(.)$ and $\emptyset(.)$ and the initial conditions for the neural networks $\hat{x}_0$ and the weights $W_{10}$ and $W_{20}$ are selected properly. The diagonal matrix A is selected Hurwitz. Also, the fractional order $q$ in (4) and step size of PC algorithm must be selected. The constants in the learning rule (23) are then selected. The larger the $k_1$ and $k_2$ parameters the faster the learning process would be, although sometimes stability is lost. We can get the system state $x_t$ from the plant and the FODNN state $\hat{x}_t$ from (4). Using the identification error $e_t$ , the weights of the FODNN are updated. Figure 5 shows the general view of the model structure used for identification in which $x_real = \begin{bmatrix} Z_p(t) \\ \frac{I_{cz}(t)}{I_p} \\ \frac{I_{eq}(t)}{I_p} \end{bmatrix}$ and inputs of the model are $u = \begin{bmatrix} 0 \\ \frac{V_{cz}(t)}{I_p} \\ \frac{V_{eq}(t)}{I_p} \end{bmatrix}$. These states and inputs have been selected for identification due to the physical model.

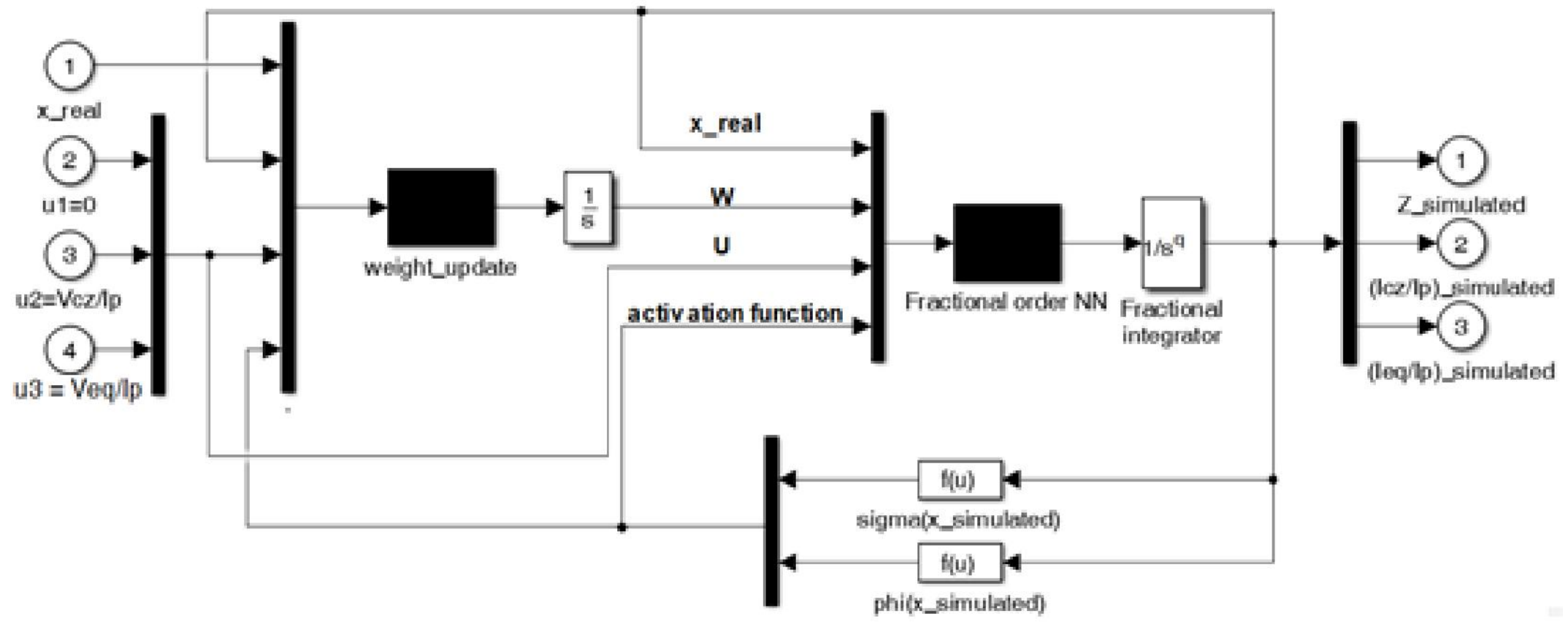

Figure 5: The simulator model structure used for identification.

To better evaluate the simulation results, the performance index, $I_\tau = \frac{1}{\tau}\int_0^\tau e^T(t)\, e(t)\, dt$ is used where $e(t)$ is the error between the real output and the identified model output. Shot (2015/27/05 Shot# 83) was used as the 'training' set and for validation of the identified fractional order model with experimental data, a shot (2015/27/05 Shot# 85) is taken with conditions similar to that of the shot used in the training step. In "validation" phase, the outputs of the plant based on the weight matrices which have been updated for the training data (2015/27/05 Shot# 83) are simulated for this shot.

The results of the identification process are presented in Figures 6-8 and Table 1-2. In Figure 6-7, the results of the identified model based on dynamic neural network, integer order dynamic neural network and FODNN have been illustrated together with the values of experimental output (2015/27/05 Shot# 83). In these figures, the plasma vertical position ($Z_p$) is zoomed for a time interval. The performance index of the shot (2015/27/05 Shot# 85), is shown in Figure 8 for both integer dynamic neural network and FODNN. As shown in Figure 8, the performance index of the three identified state variable in the shot (2015/27/05 Shot# 85) for FODNN is less than integer dynamic neural network.

In Table 1-2, the value of the performance index obtained after $t = 2\,msec$ for identification error of integer order dynamic neural network and the FODNN in training and validation phases respectively. Based on these Tables, the three identified state variables in FODNN model for the shot (2015/27/05 Shot# 83) in training step and for the shot (2015/27/05 Shot# 85) in validation step are more accurate than integer order dynamic neural network model and it is obvious that performance of the FODNN is better than the integer order neural network.

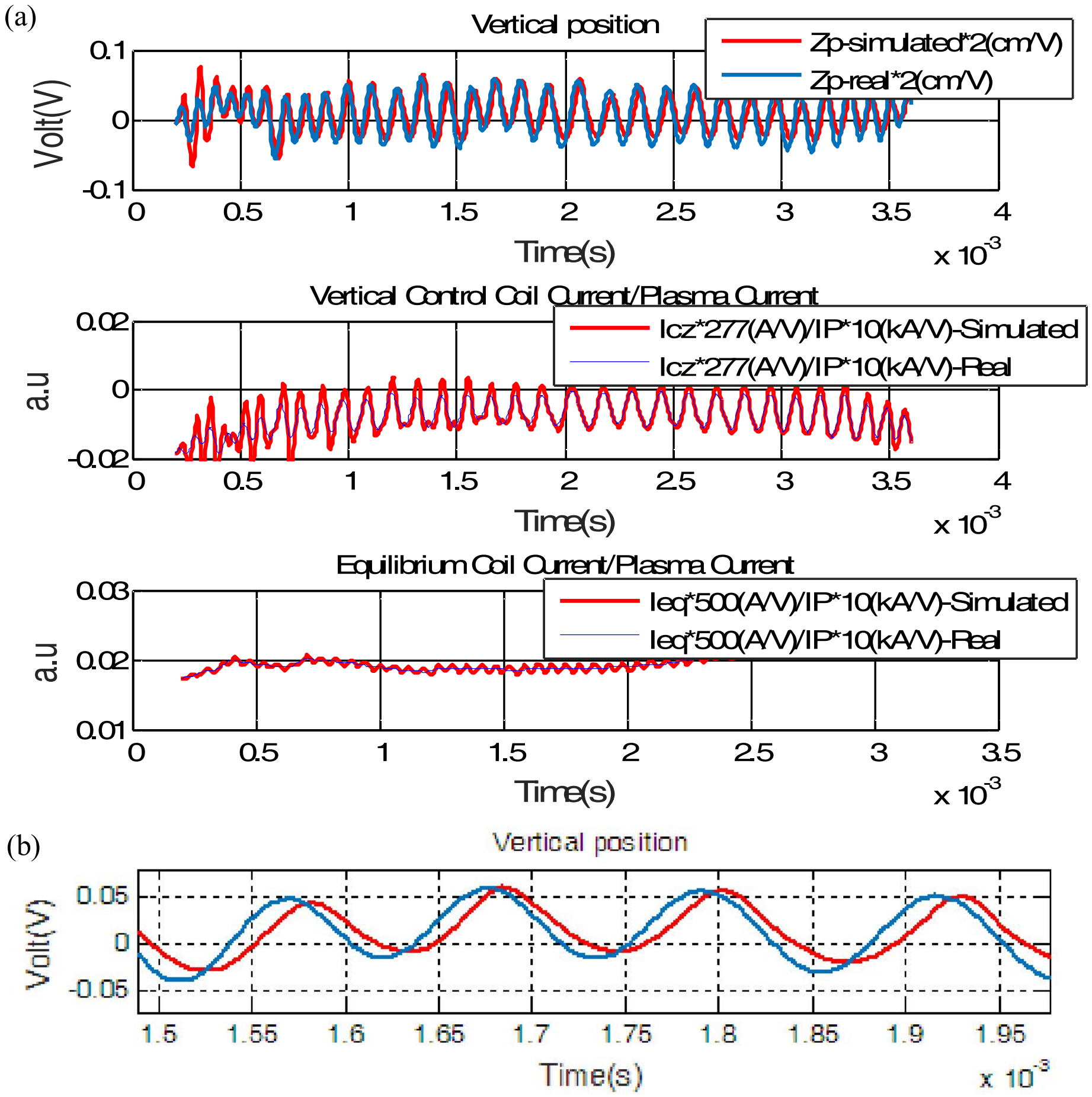

Figure. 6. (a) Experimental data for validation of the Damavand Tokamak dynamic model (shot (2015/27/05 Shot# 83)) and simulated response of a integer order dynamic neural network model, superimposed over validation.(b) Enlarged region of interest of the plasma vertical position model(shot(2015/27/05 Shot# 83))

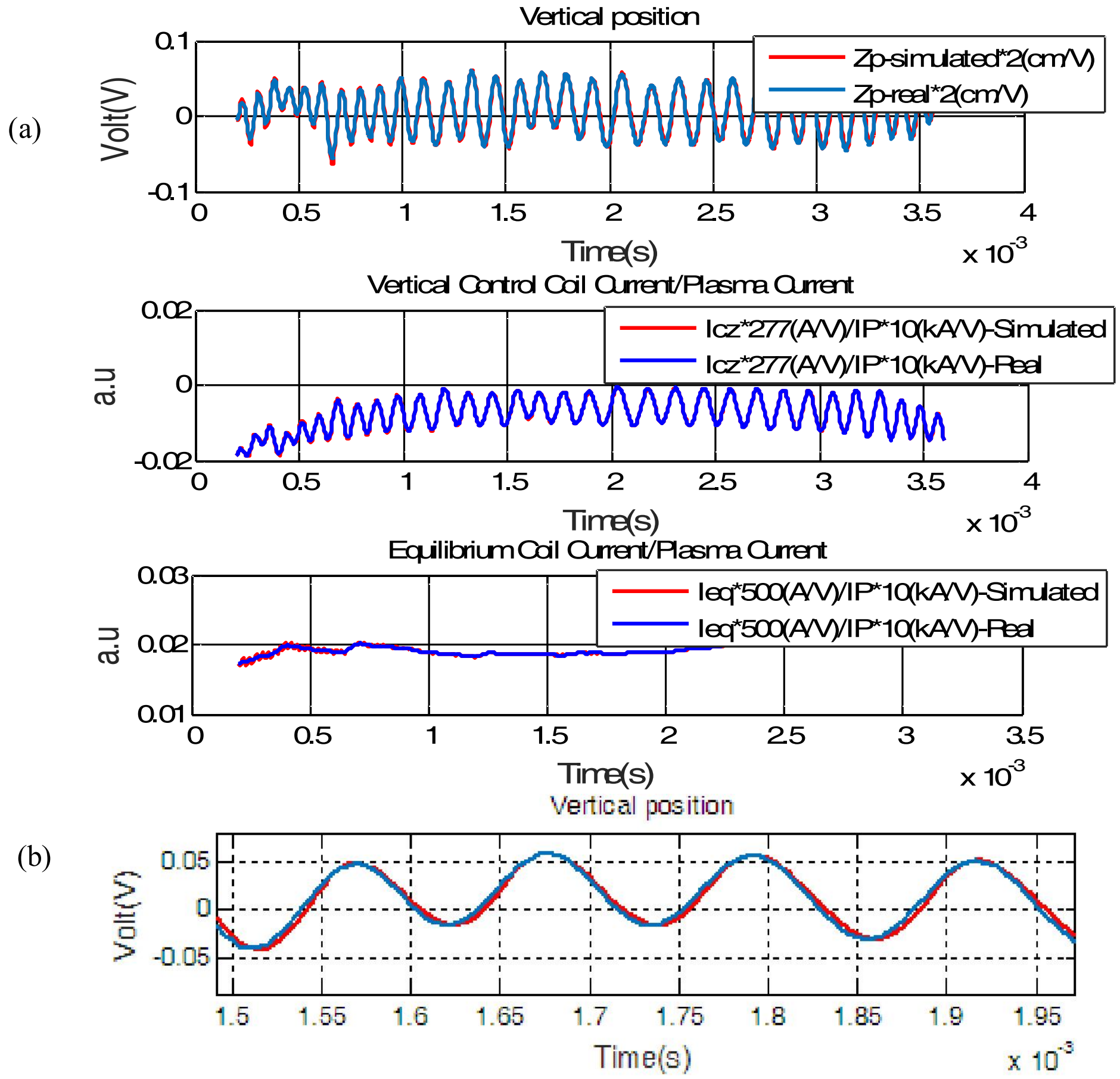

Figure. 7. (a) Experimental data for validation of the Damavand Tokamak dynamic model (shot (2015/27/05 Shot# 83)) and Simulated response of a FODNN model, superimposed over validation.(b) Enlarged region of interest of the plasma vertical position model (shot (2015/27/05 Shot# 83)).

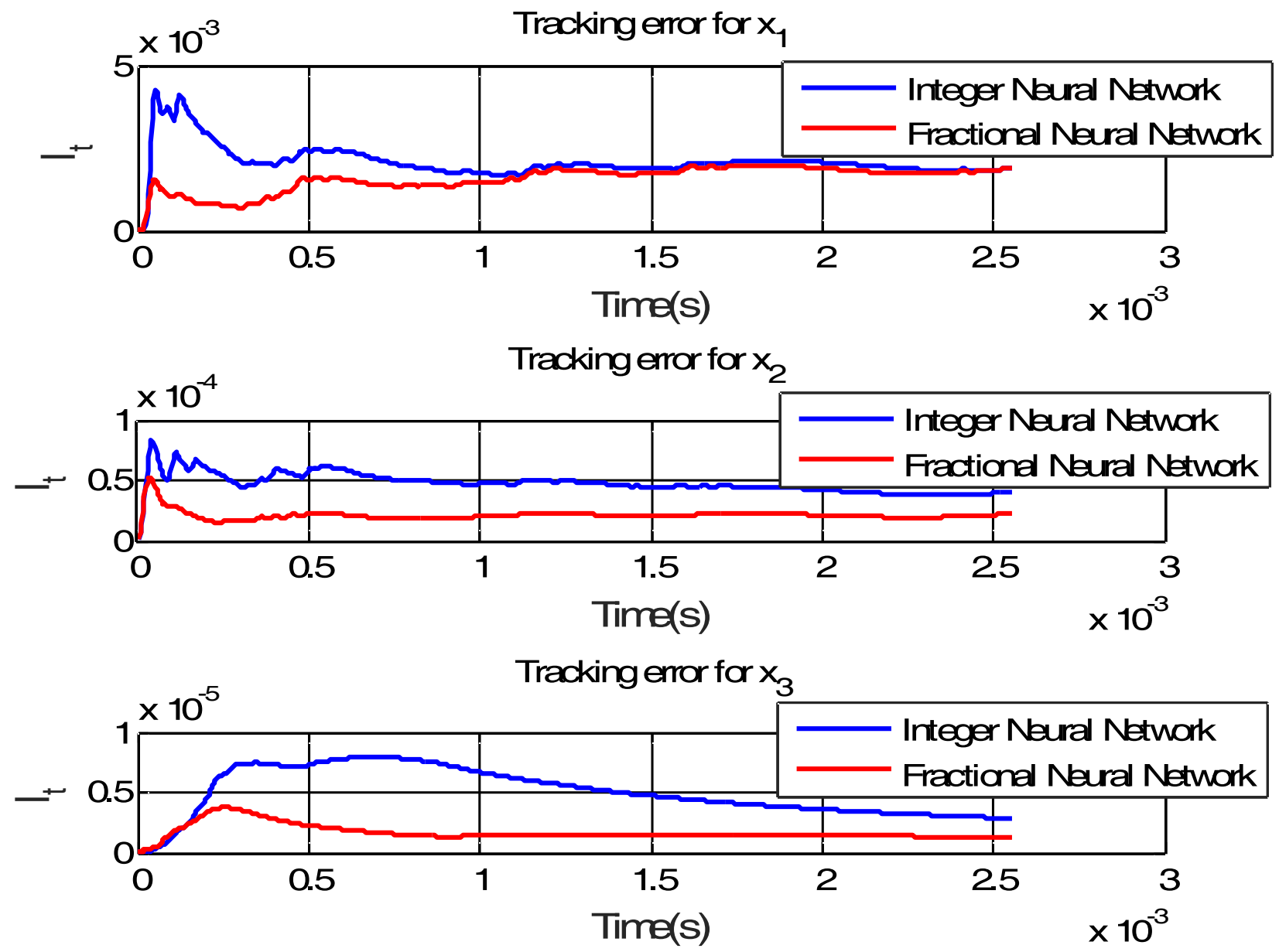

Figure 8. Performance index $I_\tau$ of integer dynamic neural network model and FODNN model (2015/27/05 Shot# 85) in validation phase.

Table 1. The performance index of three state variables in the integer order dynamic neural network model and fractional order dynamic neural network model for shot (2015/27/05 Shot# 83) in training phase.

| | **Integer Order Dynamic Neural Network** | **Fractional Order Dynamic Neural Network** |
|---|---|---|
| $I_{t=2\ msec}(e_1)$ | $39 \times 10^{-5}$ | $1.4 \times 10^{-5}$ |
| $I_{t=2\ msec}(e_2)$ | $1200 \times 10^{-8}$ | $3.9 \times 10^{-8}$ |
| $I_{t=2\ msec}(e_3)$ | $11 \times 10^{-8}$ | $2.6 \times 10^{-8}$ |

Table 2. The performance index of three state variables in the integer order dynamic neural network model and fractional order dynamic neural network model for shot (2015/27/05 Shot# 85) in validation phase.

| | **Integer Order Dynamic Neural Network** | **Fractional Order Dynamic Neural Network** |
|---|---|---|
| $I_{t=2\ msec}(e_1)$ | $20 \times 10^{-4}$ | $18 \times 10^{-4}$ |
| $I_{t=2\ msec}(e_2)$ | $4.2 \times 10^{-5}$ | $2.1 \times 10^{-5}$ |
| $I_{t=2\ msec}(e_3)$ | $3.6 \times 10^{-6}$ | $1.4 \times 10^{-6}$ |

## 5. Conclusions

This study is focused on identification of the Damavand Tokamak by using a fractional order dynamic neural network. The performance of the proposed identifier is shown by several simulation results and comparison with the experimental results. The performance index is considered to show the performance quantitatively. The ultimate boundedness of the prediction and estimation error is shown based on the Lyaponuv theory.

The performance comparison between the identified model and the integer order dynamic neural network model in different phases reveals a significant difference in their performance index. This index, for the proposed fractional order dynamic neural network is 10 times less than that of the integer order dynamic neural network in the training phase. In the validation phase, the tested data of a shot similar to the shot used in the training phase were evaluated by defined performance index. The results demonstrate the quality performance of the proposed method.

## References

[1] O. Nelles, Nonlinear System Identification: From Classical Approaches to Neural Networks and Fuzzy Models, Springer, (2001).
[2] E.O. Carew, T.C. Doehring, J.E. Barber, A.D. Freed, I. Vesely, Fractional-Order Viscoelasticity Applied to Heart Valve Tissues, Bioengineering Conference, Florida, 2003, pp. 25-29.
[3] R. Pintelon, J. Schoukens, L. Pauwels, E. Van Gheem, Diffusion systems: stability, modeling, and identification, IEEE Transactions on Instrumentation and Measurement, 54 (2005) 2061-2067.
[4] R.L. Magin, Fractional calculus models of complex dynamics in biological tissues, Computers & Mathematics with Applications, 59 (2010) 1586-1593.
[5] L. Zhang, X. Hu, Z. Wang, F. Sun, D.G. Dorrell, Fractional-order modeling and State-of-Charge estimation for ultracapacitors, Journal of Power Sources, 314 (2016) 28-34.
[6] A. Yazdizadeh, K. Khorasani, Adaptive time delay neural network structures for nonlinear system identification, Neurocomputing, 47 (2002) 207-240.
[7] G.A. Rovithakis, M.A. Christodoulou, Adaptive control of unknown plants using dynamical neural networks, IEEE Transactions on Systems, Man and Cybernetics, 24 (1994) 400-412.
[8] A. Poznyak, E. Sanchez, Nonlinear system approximation by neural networks: error stability analysis, Intelligent Automation & Soft Computing, 1 (1995) 247-257.
[9] E.N. Sanchez, A.S. Poznyak, W. Yu, Dynamic multilayer neural network for nonlinear system identification, Neural Networks, International Joint Conference on, IEEE, 1999, pp. 2109-2113.
[10] W. Yu, X. Li, Some new results on system identification with dynamic neural networks, IEEE Transactions on Neural Networks, 12 (2001) 412-417.
[11] W. Yu, X. Li, Adaptive control with multiple neural networks, American Control Conference, IEEE, 2002, pp. 1543-1548.
[12] T.T. Hartley, C.F. Lorenzo, Fractional-order system identification based on continuous order-distributions, Signal Processing, 83 (2003) 2287-2300.
[13] D. Maiti, M. Chakraborty, A. Konar, A novel approach for complete identification of dynamic fractional order systems using stochastic optimization algorithms and fractional calculus, International Conference on Electrical and Computer Engineering, 2008, pp. 867-872.
[14] A. Djouambi, A. Voda, A. Charef, Recursive prediction error identification of fractional order models, Communications in Nonlinear Science and Numerical Simulation, 17 (2012) 2517-2524.
[15] L.u. Dorcak, E.A. Gonzalez, J.a.T.´ak, J. Valsa, L. Pivka, Identification of Fractional-Order Dynamical Based on Nonlinear Function Optimization, 89 (2013).
[16] B.-M. François, S. Laurent, P. Thierry, T. Jean-Claude, Identification of Nonlinear Fractional Systems using Continuous Time Neural Networks, IFAC Workshop on Fractional Differentiation and its Applications, The Institute of Engineering of Porto (ISEP), Portugal, 2006, pp. 402-407.
[17] D. Sierociuk, I. Petras, Modeling of heat transfer process by using discrete fractional-order neural networks, 16th International Conference on Methods and Models in Automation and Robotics (MMAR), 2011, pp. 146-150.
[18] D. Sierociuk, G. Sarwas, A. Dzieliński, Discrete Fractional Order Artificial Neural Network, acta mechanica et automatica, Vol. 5, no. 2 (2011) 128-132
[19] P. Arena, L. Fortuna, D. Porto, Chaotic behavior in noninteger-order cellular neural networks, Physical Review E, 61 (2000) 776.
[20] T. Matsuzaki, M. Nakagawa, A Chaos Neuron Model with Fractional Differential Equation, Journal of the Physical Society of Japan, 72 (2003) 2678-2684.
[21] I. Petras, A Note on the Fractional-Order Cellular Neural Networks, International Joint Conference on Neural Networks, 2006, pp. 1021-1024.
[22] A. Boroomand, M. Menhaj, Fractional-Order Hopfield Neural Networks,Advances in Neuro-Information Processing, Springer Berlin Heidelberg2009, pp. 883-890.

[23] E. Kaslik, S. Sivasundaram, Nonlinear dynamics and chaos in fractional-order neural networks, Neural Networks, 32 (2012) 245-256.
[24] S. Zhang, Y. Yu, H. Wang, Mittag-Leffler stability of fractional-order Hopfield neural networks, Nonlinear Analysis: Hybrid Systems, 16 (2015) 104-121.
[25] A. Boroomand, M.B. Menhaj, On-Line nonlinear systems identification of coupled tanks via fractional differential neural networks, Control and Decision Conference, 2009. Chinese, 2009, pp. 2185-2189.
[26] A. Boroomand, M.B. Menhaj, Fractional-based approach in neural networks for identification problem, Control and Decision Conference, Chinese, 2009, pp. 2319-2322.
[27] W. Bin, A. Bruschi, O.D. Arcangelo, C. Castaldo, M.D. Angeli, L. Figini, C. Galperti, S. Garavaglia, G. Granucci, G. Grosso, S.B. Korsholm, M. Lontano, V. Mellera, D. Minelli, A. Moro, A. Nardone, S.K. Nielsen, J. Rasmussen, A. Simonetto, M. Stejner, U. Tartari, First operations with the new Collective Thomson Scattering diagnostic on the Frascati Tokamak Upgrade device, Journal of Instrumentation, 10 (2015) P10007.
[28] T. Czarski, M. Chernyshova, K.T. Pozniak, G. Kasprowicz, W. Zabolotny, P. Kolasinski, R. Krawczyk, A. Wojenski, P. Zienkiewicz, Serial data acquisition for the X-ray plasma diagnostics with selected GEM detector structures, Journal of Instrumentation, 10 (2015) P10013.
[29] H. Rasouli, C. Rasouli, A. Koohi, Identification and control of plasma vertical position using neural network in Damavand tokamak, Review of Scientific Instruments, 84 (2013) 023504.
[30] N. Darestani Farahani, F. Abbasi Davani, Experimental determination of some equilibrium parameter of Damavand tokamak by magnetic probe measurements for representing a physical model for plasma vertical movement, Review of Scientific Instruments, 86 (2015) 103510.
[31] N.D. Farahani, F.A. Davani, Simulation of open-loop plasma vertical movement response in the Damavand tokamak using closed-loop subspace system identification, Journal of Instrumentation, 11 (2016) P02006.
[32] D. del-Castillo-Negrete, B.A. Carreras, V.E. Lynch, Nondiffusive Transport in Plasma Turbulence: A Fractional Diffusion Approach, Physical Review Letters, 94 (2005) 065003.
[33] H. Rasouli, A. Fatehi, Design of set-point weighting PIλ+ Dµ controller for vertical magnetic flux controller in Damavand tokamak, Review of Scientific Instruments, 85 (2014) 123508.
[34] H. Rasouli, A. Fatehi, H. Zamanian, Design and implementation of fractional order pole placement controller to control the magnetic flux in Damavand tokamak, Review of Scientific Instruments, 86 (2015) 033503.
[35] H. Rasouli, C. Rasouli, A. Koohi, Identification and control of plasma vertical position using neural network in Damavand tokamak, Review of Scientific Instruments, 84 (2013).
[36] N. Darestani Farahani, F. Abbasi Davani, Experimental Determination of Some Equilibrium Parameter of Damavand tokamak by Magnetic Probe Measurements for Representing a physical Model for Plasma Vertical Movement, Review of Scientific Instruments, 86 (2015).
[37] I. Podlubny, Fractional Differential Equations. An Introduction to Fractional Derivatives, Fractional Differential Equations, Some Methods of Their Solution and Some of Their Applications, Academic Press, San Diego - New York - London1999.
[38] N. Darestani Farahani, Damavand tokamak plasma equilibrium modeling for preparations of desigening linear time-invariant model-based controller, Radiation application department, Shahid beheshti university, Iran, Tehran, 2016, pp. 191.